\documentclass[conference]{IEEEtran}
\IEEEoverridecommandlockouts
\usepackage{cite}
\usepackage{graphicx}
\usepackage{epstopdf}
\usepackage{amsmath,amssymb,amsfonts}
\usepackage{algorithmic}
\usepackage{textcomp}
\usepackage{xcolor}
\usepackage{tikz}
\usetikzlibrary{arrows.meta, positioning, decorations.markings}
\usepackage{tabularx,booktabs}
\usepackage[flushleft]{threeparttable}

\usepackage{booktabs,ragged2e}
\usepackage{multirow}
\usepackage{subfigure}

\usepackage{bm}
\usepackage[colorlinks,
            linkcolor=blue,
            anchorcolor=blue,
            citecolor=blue
            ]{hyperref}

\def\BibTeX{{\rm B\kern-.05em{\sc i\kern-.025em b}\kern-.08em
    T\kern-.1667em\lower.7ex\hbox{E}\kern-.125emX}}

\begin{document}

\title{Secure Semantic Communication for Image Transmission in the Presence of Eavesdroppers

}
\author{\IEEEauthorblockN{Shunpu Tang\IEEEauthorrefmark{2},   
Chen Liu\IEEEauthorrefmark{4},
    Qianqian Yang\IEEEauthorrefmark{2}\IEEEauthorrefmark{1},
    Shibo He\IEEEauthorrefmark{4},
    Dusit Niyato\IEEEauthorrefmark{8}
    }
\IEEEauthorblockA{ \IEEEauthorrefmark{2}College of Information Science and Electronic Engineering, Zhejiang University, Hangzhou, China \\
\IEEEauthorrefmark{4}College of Control Science and Engineering, Zhejiang University, Hangzhou, China \\
\IEEEauthorrefmark{8} School of Computer Science and Engineering, Nanyang Technological University, Singapore \\
Email: 
\{tangshunpu, liu777ch, qianqianyang20, s18he\}@zju.edu.cn, 
dniyato@ntu.edu.cn
\vspace{-5mm}
}}

%


\maketitle

\begin{abstract}
Semantic communication (SemCom) has emerged as a key technology for the forthcoming sixth-generation (6G) network, attributed to its enhanced communication efficiency and robustness against channel noise. However, the open nature of wireless channels renders them vulnerable to eavesdropping, posing a serious threat to privacy. To address this issue, we propose a novel secure semantic communication (SemCom) approach for image transmission, which integrates steganography technology to conceal private information within non-private images (host images). Specifically, we propose an invertible neural network (INN)-based signal steganography approach, which embeds channel input signals of a private image into those of a host image before transmission. This ensures that the original private image can be reconstructed from the received signals at the legitimate receiver, while the eavesdropper can only decode the information of the host image. Simulation results demonstrate that the proposed approach maintains comparable reconstruction quality of both host and private images at the legitimate receiver, compared to scenarios without any secure mechanisms. Experiments also show that the eavesdropper is only able to reconstruct host images, showcasing the enhanced security provided by our approach.

\end{abstract}


\section{Introduction}

Recently, semantic communication (SemCom) has emerged as a key technology for the forthcoming sixth-generation (6G) network and has attracted significant research interest from both academia and industry. Unlike conventional digital communication systems that transmit data without considering its underlying meaning, SemCom prioritizes the transmission of task-related semantic information, thereby significantly enhancing communication efficiency\cite{Semantic1}. This enables SemCom to closely align with the requirements of innovative applications envisioned for the 6G network, including smart cities, autopilot systems, extended reality (XR), artificial intelligence generated content (AIGC), and metaverse\cite{semantic_IOT,AIGC_semantic}. 

With the advancement of deep learning (DL) technology, the authors in \cite{DeepJSCC} proposed a DL-based SemCom approach for wireless image transmission, achieving superior image reconstruction quality compared to conventional digital communication systems. Following this, lots of works have enhanced the performance of SemCom systems\cite{chen2023commin,tianxiao_GAN,han2022semantic}. 
However, privacy leakage may occur more easily in SemCom systems due to its intrinsic characteristics\cite{Semantic_security_zhaohui,Semantic_security_in_IOT}. Specifically, the open nature of the wireless channel enables eavesdroppers to capture the transmitted signals. If eavesdroppers gain access to the semantic decoder, they may decode some or all of the private or sensitive semantic information, even when their channel conditions are significantly worse than those of legitimate receivers.
Classical physical layer security (PLS) methods for conventional digital communication systems typically focus on improving secure channel capacity or reducing secure outage probability (SOP). However, these methods are not directly applicable to SemCom systems, which typically operate in a joint source-channel coding manner that directly maps the source data to complex channel symbols without quantization.

To address this issue, researchers have developed secure SemCom approaches which reduce the decoding ability of eavesdroppers to protect privacy. For instance, in \cite{Semantic_security_maojun}, the authors introduced a novel secure mean square errors (MSE) loss function to train the semantic encoder and decoder. This ensures that the semantic decoder fails to decode the signal successfully when the channel conditions are poorer than those during the training stage, thereby preventing eavesdroppers from exploiting the semantic decoder to overhear, as the channel conditions of eavesdropping link are typically worse. However, this approach may hurt the robustness of SemCom system when faces different channel conditions. In \cite{Semantic_security_yuhao}, the authors proposed to perform random permutation and substitution on the output of the semantic encoder before transmission, which can defend model inversion attacks (MIA) from eavesdroppers. 

Recent studies \cite{DeepJSCC_Encryption, SemCom_Encrypted,vis_protect} have integrated encryption schemes into SemCom to protect transmitted semantic information from eavesdropping. In \cite{DeepJSCC_Encryption}, the authors proposed incorporating quantization operation and a public-key encryption scheme into SemCom, achieving superior performance compared to digital communication systems with advanced encryption standards (AES) encryption. Similarly, the authors in \cite{SemCom_Encrypted} introduced an adversarial encryption training scheme to protect the transmitted semantic information. The authors in \cite{vis_protect} proposed a novel vision protection approach, where a neural network (NN)-based protection module and a deprotection module were designed to realize transmitting semantic information in a protected domain. Additionally, in \cite{Semantic_security_weixaun}, the authors utilized a superposition code to maximize the symbol error probability (SER) of eavesdroppers without impacting the main link. While these approaches effectively reduce the risk of leakage, 
it is worth noting that these approaches will result in over-distortion in the images decoded by eavesdroppers, which may raise suspicions from eavesdroppers and prompt them to exert jamming attacks to corrupt communication.

In this paper, inspired by the concept of covert communication, we explore secure SemCom for private image transmission from a different perspective compared to the existing studies. In particular, our goal is to mislead potential eavesdroppers by leveraging steganography technology to conceal the private image within any non-sensitive image, referred to as the \textit{host image}. Specifically, steganography technology works by generating an image, referred to as the \textit{container image}, which conceals privacy-sensitive content while preserving visual perceptual characteristics similar to those of the host image. Then, the sender transmits the \textit{container image}, and eavesdroppers can only decode the semantic information of the host image, thereby enhancing the security of the SemCom and preventing privacy leakage to eavesdroppers. The main contributions of this paper can be summarized as follows:
\begin{itemize}
    \item We propose a novel secure SemCom approach for the transmission of private images that employs steganography to hide these privacy-sensitive images within host images. This approach ensures that potential eavesdroppers can only decode semantic information related to the host image, effectively avoiding suspicion from eavesdroppers.
    \item We propose an invertible neural network (INN) based signal steganography to apply steganography on the channel input signal before transmission, instead of directly on the source images. This approach effectively improves the performance of steganography and image reconstruction in SemCom. 
    \item We conduct simulations on the CelebAmasked-HQ dataset to verify the effectiveness of the proposed approach. The numerical results show that the proposed approach maintains comparable reconstruction quality of the private image at the legitimate receiver compared to the scenario without any secure mechanisms, while the eavesdropper is only able to reconstruct host images. This validates the enhanced security provided by our approach.
\end{itemize}
\vspace{-1mm}
\begin{figure}
    \centering
    \includegraphics[width=\linewidth]{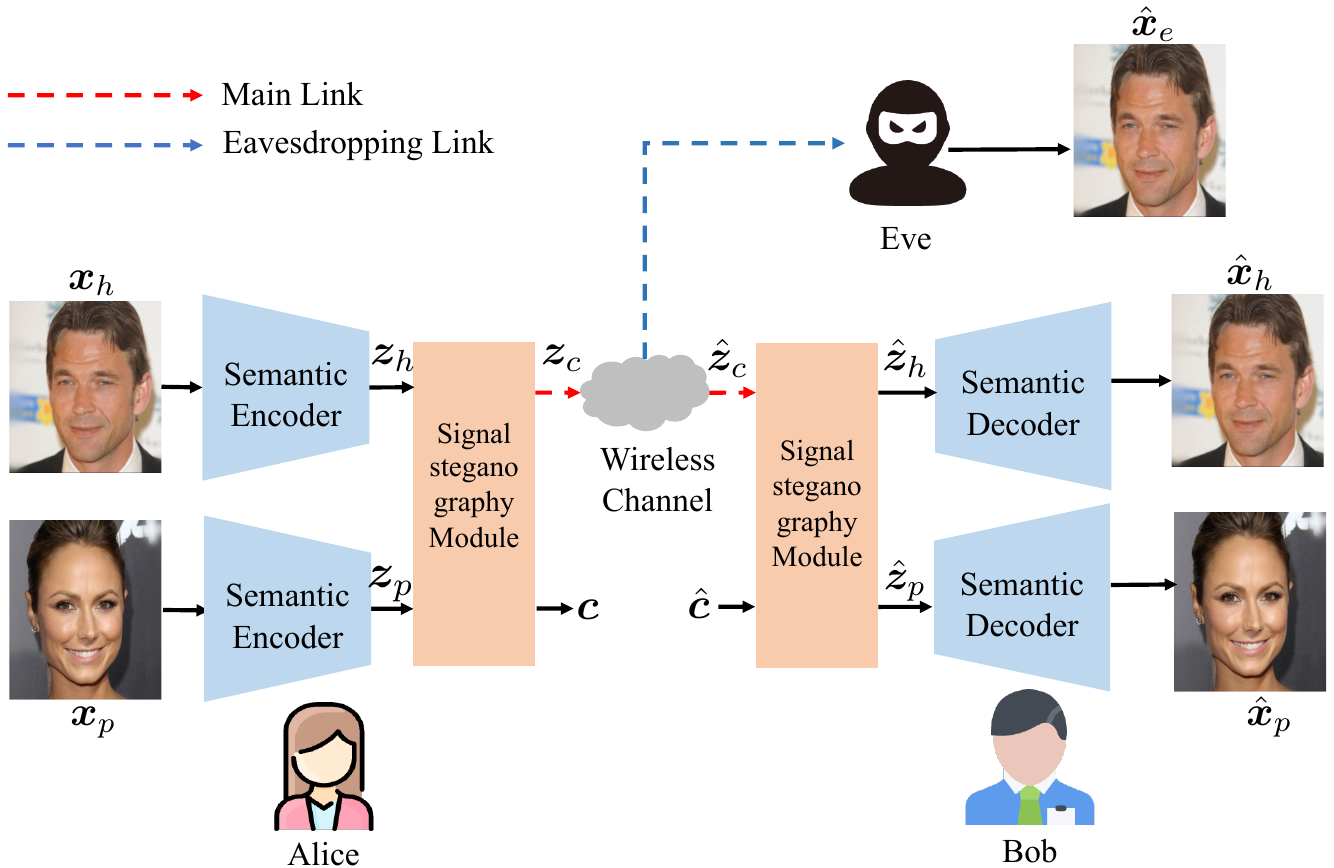}
    \caption{Overview of the proposed secure SemCom with signal steganography. }
    \label{fig:Illustraion}

\end{figure}
\begin{figure*}[!t]
    \centering
    \includegraphics[width=0.65\linewidth]{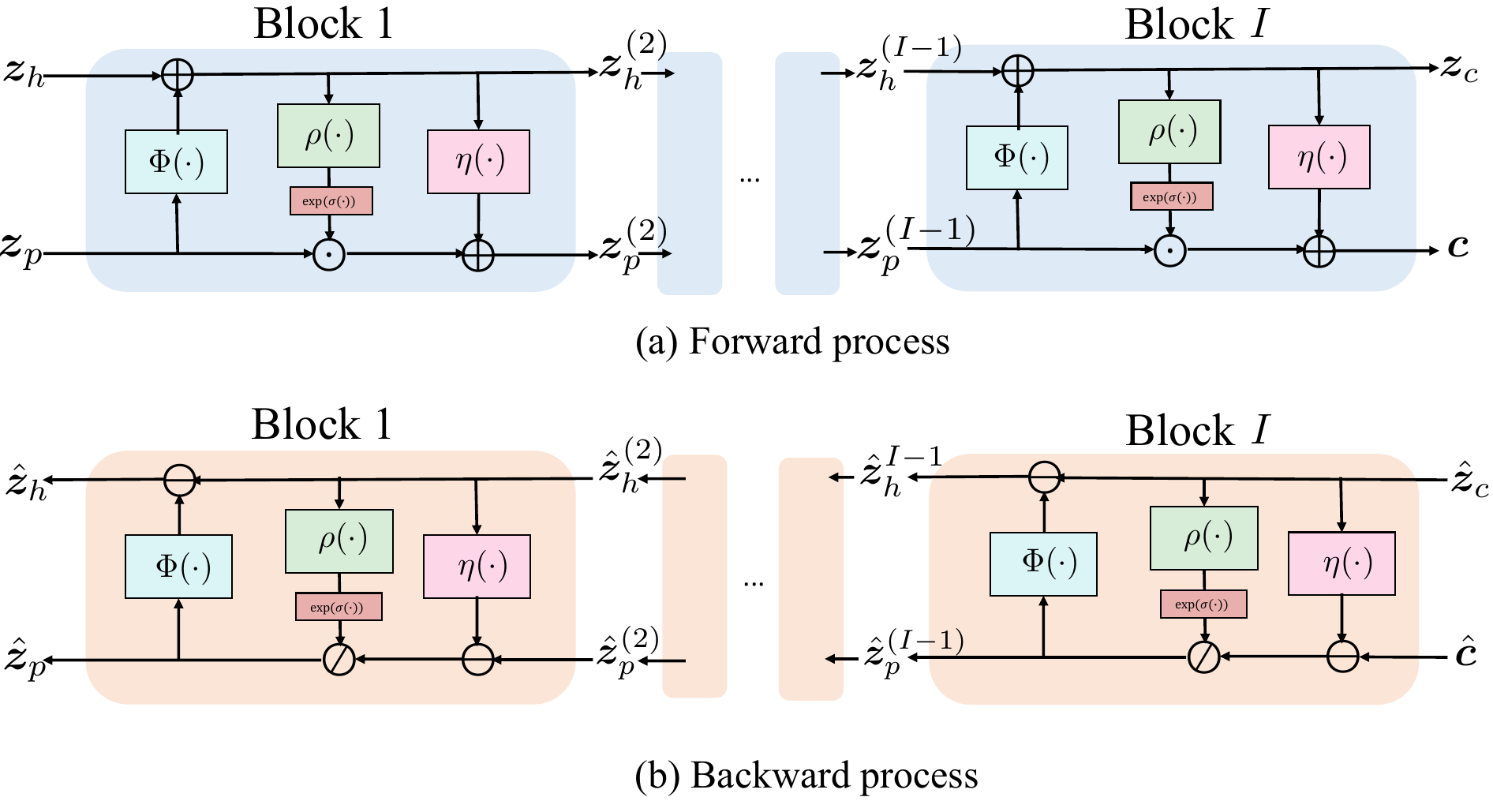}
    \caption{Architecture of the INN-based signal steganography module, which consists of several invertible blocks.}
    \label{fig:INN}
 
 \end{figure*}
\section{Problem Statement}
We consider the problem of transmitting images over a noisy channel in the presence of a passive eavesdropper, who only attempts to overhear the transmitted signal but does not send any signal. We first present the system model of SemCom for image transmission in this section and then introduce two eavesdropping methods for SemCom: naive decoding and model inversion attack. 
\subsection{SemCom for Image Transmission}
We consider a SemCom system for image transmission, which consists of a semantic encoder at the transmitter and a semantic decoder at the receiver. Denote a source image by $\bm{x} \in \mathbb{R}^{N_C \times N_H \times N_W}$. $N_C$, $N_H$, and $N_W$ represent the number of channels, height, and width of the RGB image, respectively, and $N=N_C \times N_H \times N_W$ is the source bandwidth. The semantic encoder, denoted by $\mathcal{E}(\cdot)$, maps the source image $\bm{x}$ to the $k$-dim complex channel input signal $\bm{z}\in \mathbb{C}^{k}$ with normalized power, as given by
\begin{equation}
\bm{z} = \mathcal{E}(\bm{x}; \theta_1),
\end{equation}
where $\theta_1$ is the parameters of the semantic encoder. We have the bandwidth compression ratio to be $k/N$. The signal $\bm{z}$ is then transmitted over a noisy communication channel. The received signal at the receiver is given by
\begin{equation}
    \hat{\bm{z}}=\bm{z}+\bm{n},
\end{equation}
where $\bm{n}$ denotes additive white noise following a complex Gaussian distribution $\mathcal{CN}(0,\sigma^2\bm{I})$. The receiver utilizes a semantic decoder $\mathcal{D}(\cdot)$ to reconstruct the original image $\bm{\hat{x}}$ from $\hat{\bm{z}}$, expressed as
\begin{equation}
\hat{\bm{x}} = \mathcal{D}(\hat{\bm{z}}; \theta_2),
\end{equation}
where $\theta_2$ represents the parameters of the semantic decoder. The parameters $\theta_1$ and $\theta_2$ are trained by minimizing the mean squared error (MSE) loss function using the gradient descent algorithm, given by
\begin{equation}
    \min_{\{\theta_1, \theta_2\}} \mathcal{L}(\bm{x},\hat{\bm{x}})=||\bm{x}-\hat{\bm{x}}||^2_2.
\end{equation}
\subsection{Eavesdropping in SemCom}
We consider a scenario where a passive eavesdropper, named Eve, aims to intercept private information from a transmitted signal $\hat{\bm{z}}$. The received signal at Eve's end can be expressed as
\begin{equation}
    \hat{\bm{z}}_e=\bm{z}+\bm{n_e},
\end{equation}
where $\bm{n_e}$ represents the additive white noise of the eavesdropping channel, following a complex Gaussian distribution $\mathcal{CN}(0,\sigma_e^2\bm{I})$. We outline two typical strategies that Eve can employ to obtain private information in the following.
\subsubsection{Naive Decoding}
The eavesdropper can access the semantic decoder $\mathcal{D}$ of the SemCom model, which can be utilized to reconstruct the image $\hat{\bm{x}}_e$  from the received signal $\hat{\bm{z}}_e$, given by
\begin{equation}
\hat{\bm{x}}_e = \mathcal{D}(\hat{\bm{z}}_e; \theta_2),
\end{equation}
where we recall that $\theta_2$ represents the parameters of the semantic decoder.
\subsubsection{Model Inversion Attack (MIA)}
In this scenario, Eve lacks prior knowledge of the semantic decoder but possesses information about the semantic encoder. Hence, Eve attempts to reverse-engineer the semantic encoder to extract private information\cite{Semantic_security_yuhao}. This approach, known as the model inversion attack, is formulated as an optimization problem:
\begin{equation}
\hat{\bm{x}}_e = \arg\min_{\bm{y}} \mathcal{L}\bigg (\hat{\bm{z}}_e,\mathcal{E}(\bm{y};\theta_1)\bigg ),
\end{equation}
where we recall that $\mathcal{E}$ denotes the semantic encoder with parameters $\theta_1$, and $\mathcal{L}$ represents the loss function evaluating the distortion between two images.


\begin{figure*}[!t]
    \centering
    \subfigure[]{\label{fig:Bob_a}\includegraphics[width=0.32\linewidth]{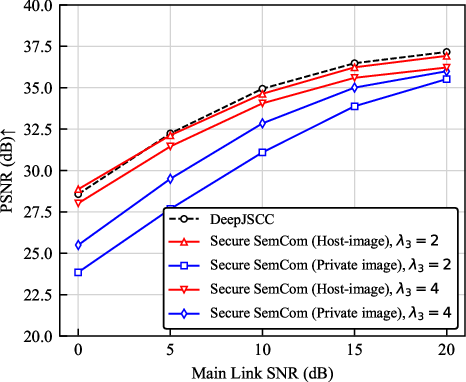}}
    \hfill
    \subfigure[]{\label{fig:Bob_b}\includegraphics[width=0.32\linewidth]{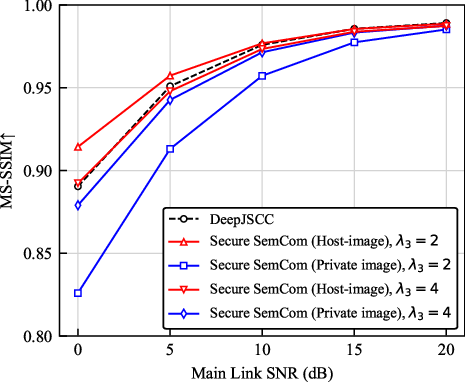}}
    \hfill
    \subfigure[]{\label{fig:Bob_c}\includegraphics[width=0.32\linewidth]{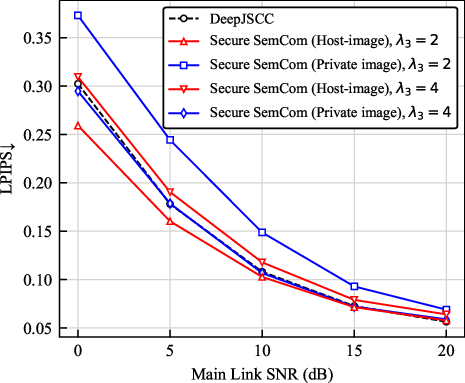}}
    \caption{PSNR and MS-SSIM of reconstructed images at Bob by DeepJSCC and the proposed secure SemCom approach, where the main link SNR varies from 0 to 20 dB. 
}
\end{figure*}
\section{Proposed approach}
In this section, we propose a secure SemCom approach to enhance the security of image transmission. In particular, we employ steganography to embed a private image within a non-sensitive host image to deceive Eve. A straightforward way is to directly exert steganography at the source images. However, this might pose challenges for the well-trained SemCom system to extract and transmit semantic information from the container image, and hence deteriorate the reconstruction performance of the private image at the legitimate 
 receiver. Therefore, we propose to exert steganography at the level of channel input signals instead, which is introduced in detail in the following.

\subsection{Secure SemCom with Signal Steganography}

As shown in \autoref{fig:Illustraion}, the proposed secure SemCom system consists of a pre-trained semantic encoder and semantic decoder at the legitimate sender and receiver, Alice and Bob, respectively. Additionally, we introduce an INN-based signal steganography module at both Alice's and Bob's ends, which embeds the channel input signal of the private image into that of the host image.

Mathematically, we denote the host image and private image as $\bm{x}_h$ and $\bm{x}_p$, respectively. These two images are input into the semantic encoder to obtain the corresponding channel input signals $\bm{z}_h$ and $\bm{z}_p$, as given by
\begin{equation}
    \begin{aligned}
        \bm{z}_h = \mathcal{E}(\bm{x}_h; \theta_1), \\
         \quad \bm{z}_p = \mathcal{E}(\bm{x}_p; \theta_1).
    \end{aligned}
\end{equation}
Then, steganography is applied to these channel input signals to obtain the channel input signal $\bm{z}_c$ of the container image, given by
\begin{equation} 
(\bm{z}_{c}, \bm{l}) = \mathcal{S}(\bm{z}_h,\bm{z}_p; \phi),
\end{equation}
where $\mathcal{S}(\cdot)$ represents the INN-based steganography module parameterized by $\phi$, and $\bm{l}$ denotes the information lost during this steganography operation. \textcolor{black}{We note that the dimension of $\bm{z}_c$ is identical to that of $\bm{z}_p$ and $\bm{z}_h$.}
The channel input signal $\bm{z}_c$ is then transmitted over a noisy communication channel, with $\hat{\bm{z}}_c$ and $\hat{\bm{z}}_{c,e}$ denoting the received signals at Bob and Eve, respectively. At Bob, the reverse operation of the same INN-based steganography module is utilized to reconstruct the original channel input signals, given by
\begin{equation}
        (\hat{\bm{z}}_h, \hat{\bm{z}}_p)= \mathcal{S}^{-1}(\hat{\bm{z}}_c, \hat{\bm{l}}; \phi),
\end{equation}
where $\hat{\bm{l}}$ is the estimated lost information at Bob, which is either a predefined constant value or sampled from a given distribution. Subsequently, Bob can reconstruct both the host and private images, $\hat{\bm{x}}_h$ and $\hat{\bm{x}}_p$, respectively, using the semantic decoder. In contrast, Eve can only employ naive decoding or MIA on the received signal $\hat{\bm{z}}_{c,e}$.

\subsection{INN-based Signal Steganography Module}
To obtain the channel input signal $\bm{z}_c$, we employ the INN to apply steganography on the channel input signals $\bm{z}_h$ and $\bm{z}_p$. INN enables reversible operations, ensuring precise reconstruction of the original signals from the INN output signals. Utilizing INN for signal steganography effectively mitigates introduced errors, ensuring the successful reconstruction of the private image by the semantic decoder at Bob.

The architecture of INN, as shown in \autoref{fig:INN}, comprises multiple invertible blocks. Following the design of the invertible block in \cite{ISN}, we utilize additive affine transformations. Let $\bm{z}_h^i$
and $\bm{z}_p^i$ denote the input of the $i$-th invertible block, and $\bm{z}_h^{(i+1)}$ and $\bm{z}_p^{(i+1)}$ denote the output of the $i$-th invertible block. The forward operation of the $i$-th invertible block is given by
\begin{equation}
    \begin{aligned}
        \bm{z}_h^{(i+1)} &= \bm{z}_{h}^{(i)} + \Phi(\bm{z}_{p}^{(i)}), \\
        \bm{z}_p^{(i+1)} &= \bm{z}_{p}^{(i)} \odot \exp\bigg (\rho(\bm{z}_h^{(i+1)})\bigg)+\eta(\bm{z}_h^{(i+1)}),
    \end{aligned}
\end{equation}
where $\Phi(\cdot)$, $\rho(\cdot)$, and $\eta(\cdot)$ represent arbitrary additive affine transformations, realized using convolutional blocks in practice, and $\odot$ denotes the Hadamard product. 
After $I$ invertible blocks, the container channel input signal  $\bm{z}_c=\bm{z}_h^{(I)}$ and the lost information $\bm{l}=\bm{z}_p^{(I)}$ are obtained.

For the backward operation of INN, the original channel input signals $\hat{\bm{z}}_h$ and $\hat{\bm{z}}_p$ are reconstructed from the received signal $\hat{\bm{z}}_c$ and the estimated lost information $\hat{\bm{l}}$. Initializing $\hat{\bm{z}}_h^{(i)}=\hat{\bm{z}}_c$ and $\hat{\bm{z}}_p^{(i)}=\hat{\bm{l}}$, the backward operation of the invertible blocks is iteratively applied, given by
\begin{equation}
    \begin{aligned}
        \hat{\bm{z}}_p^{(i)} &= \bigg (\hat{\bm{z}}_p^{(i+1)}-\eta(\hat{\bm{z}}_h^{(i+1)})\bigg)\odot \exp\bigg (-\rho(\hat{\bm{z}}_h^{(i+1)})\bigg), \\
        \hat{\bm{z}}_h^{(i)} &= \hat{\bm{z}}_h^{(i+1)} - \Phi(\hat{\bm{z}}_p^{(i)}).
    \end{aligned}
\end{equation}
We note that only $\bm{z}_c$ is transmitted over the wireless channel. $\hat{\bm{l}}$ is a predefined constant value or sampled from a given distribution to approximate the lost information $\bm{l}$.

\begin{table*}
    \centering
    \caption{Performance comparison in terms of MS-SSIM and LPIPS, for DeepJSCC and the proposed secure SemCom approach at Eve, where SNR of the eavesdropping link is set to 5dB.}
    \begin{threeparttable}
        
    \begin{tabularx}{0.6\linewidth}{lcccc}
    \toprule
    \multirow{4}{*}{Method} & \multicolumn{2}{c}{Naive Decoding} & \multicolumn{2}{c}{MIA} \\
    \cmidrule(lr){2-3} \cmidrule(lr){4-5} 
    & MS-SSIM & LPIPS & MS-SSIM & LPIPS\\
    \midrule
    DeepJSCC (Private image) &  0.951& 0.178 & 0.604 & 0.627 \\
    \midrule
    Secure SemCom (Private image), $\lambda_3$=2  & 0.275 & 0.639 & 0.166 & 0.737 \\
    Secure SemCom (Private image), $\lambda_3$=4 &  0.270 & 0.639 & 0.160 & 0.744 \\
    \midrule
    Secure SemCom (host image), $\lambda_3$=2 & 0.943 & 0.195 & 0.686 & 0.584 \\
    Secure SemCom (host image), $\lambda_3$=4 &  0.936& 0.212 & 0.663 & 0.602 \\
    \bottomrule
    \end{tabularx}
    \end{threeparttable}
    \label{tab:Eve}  
\end{table*}
\subsection{Training procedure and Loss function}
To mitigate the risk of privacy leakage, we design a training procedure and loss function for the proposed signal steganography module. The first objective of this training is to minimize the difference between $\bm{z}_c$ and $\bm{z}_{h}$ in order to hide the private information $\bm{z}_{p}$ as much as possible. Since $\bm{l}$ is not transmitted to Bob, it is also necessary to minimize the difference between $\bm{l}$ and $\hat{\bm{l}}$, which is either a given constant or sampled from a given distribution. This ensures that Bob can successfully reconstruct the original channel input signals $\bm{z}_h$ and $\bm{z}_p$ through the backward process of the INN. Therefore, the loss function of the forward process of INN can be expressed as
\begin{equation}
    \mathcal{L}_{\text{forward}} = \lambda_1||\bm{z}_c-\bm{z}_h||^2_2+\lambda_2||\bm{l}-\hat{\bm{l}}||^2_2,
\end{equation}
where $\lambda_1>0$ and $\lambda_2>0$ are hyperparameters to balance these two terms. At Bob's end, the objective is to reconstruct the input signals of INN, $\bm{z}_p$ and $\bm{z}_h$. Hence, we have the loss function of the backward process given by
\begin{equation}
    \mathcal{L}_{\text{backward}} = \lambda_3||\bm{z}_p-\hat{\bm{z}}_p||^2_2+\lambda_4||\bm{z}_h-\hat{\bm{z}}_h||^2_2,
\end{equation}
where $\lambda_3>0$ and $\lambda_4>$ are the hyperparameters to balance these two terms. Moreover, to ensure the reconstruction performance of the private image at Bob, we also include a loss term to minimize the difference between $\bm{x}_p$ and $\hat{\bm{x}}_p$, given by
\begin{equation}
    \mathcal{L}_{\text{privacy}} = \lambda_5||\bm{x}_p-\hat{\bm{x}}_p||^2_2, 
\end{equation}
where $\lambda_5>0$ is used to control the weight of this loss. Finally, the overall loss function of the proposed approach is
\begin{equation}
    \mathcal{L}_{total} = \mathcal{L}_{\text{forward}}+\mathcal{L}_{\text{backward}}+\mathcal{L}_{\text{privacy}}.
\end{equation}
The total loss is computed after executing the forward and backward operation of INN taking into account the Gaussian noisy channel, following which we update the parameters of the INN-based signal steganography module using the gradient descent method. We note that this training process can be conducted without any prior knowledge about Eve.

\section{Simulation}
In this section, we present the simulation results to verify the effectiveness of the proposed approach. We first introduce the implementation details and then present the simulation results and discussions.

\subsection{Implementation Details}
\vspace{-2mm}
We use the well-known CelebAmasked-HQ dataset in the simulation, which contains 25,000 images for training and 5000 images for testing. We employ the semantic encoder and semantic decoder from \cite{DeepJSCC} with a bandwidth compression ratio (BCR) of 1/12, which are pre-trained on the entire training set with uniformly distributed channel SNR ranging from 0 to 20 dB. Since the steganography module does not change the output dimension, the BCR of the proposed SemCom stays 1/12. Additionally, 2,500 pairs of host and private images, randomly sampled from the training set, are used to train the signal steganography module with uniformly distributed channel SNR ranging from 0 to 20 dB. During the training stage, the image patch size is set to $256\times 256$, while during testing, it is adjusted to $512\times 512$. The INN comprises 8 invertible blocks. For the training of INN, we use the Adam optimizer with a learning rate of $3 \times 10^{-4}$ and a batch size of 128. Values for hyperparameters $\lambda_1$, $\lambda_2$, $\lambda_3$, $\lambda_4$, and $\lambda_5$ are set to 1, 2, 2, 1, and 1, respectively. We compare the performance of the proposed approach with conventional DeepJSCC, which transmits private images without employing any secure mechanisms. We evaluate performance in terms of peak signal-to-noise ratio (PSNR), multi-scaled structural similarity index (MS-SSIM) and learned perceptual image patch similarity (LPIPS).

 \subsection{Reconstruction performance at Bob}

 \begin{figure}[!t]
    \centering
    \subfigure[Naive Decoding]{\label{fig:Eve_a}\includegraphics[width=0.49\linewidth]{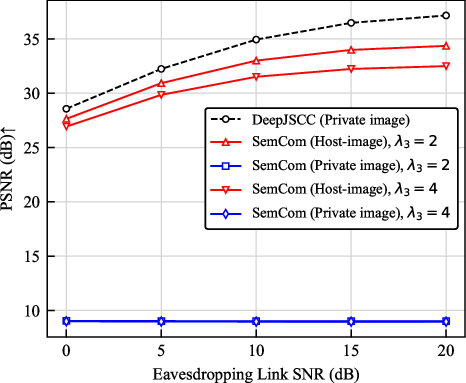}}
    \subfigure[MIA]{\label{fig:Eve_b}\includegraphics[width=0.49\linewidth]{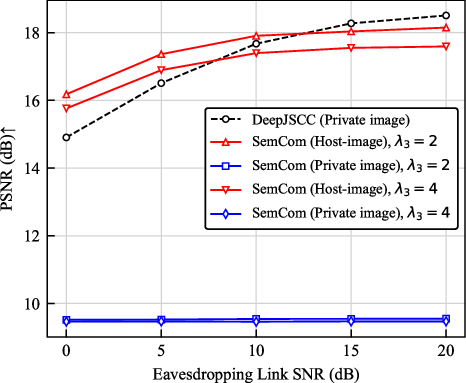}}
    \caption{PSNR of reconstructed images at Eve by DeepJSCC and the proposed secure SemCom approach with naive decoding and MIA eavesdropping, respectively, where SNR of the eavesdropping link varies from 0 to 20 dB. \vspace{-1mm}}
\end{figure}

In \autoref{fig:Bob_a} and \autoref{fig:Bob_a}, we present the PSNR, MS-SSIM and LPIPS results comparing the reconstructed host and private images by Bob with the original ones, while varying the SNR of the main link from 0 to 20 dB and setting $\lambda_3$ to 2 and 4, respectively. We also include the results of DeepJSCC, which directly transmits images, for comparison. From this figure, we observe that the proposed approach maintains comparable reconstruction quality of the private image compared to the scenario without any secure mechanisms. Notably, for the host image, the proposed approach even outperforms DeepJSCC in terms of MS-SSIM and LPIPS when the main link SNR is less than 10 dB and $\lambda_3=4$. This is because INN enables invertible operations, which helps mitigate the impact of channel noise. Moreover, the MS-SSIM and LPIPS results of the private image are also comparable to DeepJSCC, demonstrating the effectiveness of the proposed approach. We observe that adjusting the value of $\lambda_3$ allows for a trade-off between the reconstruction performance of the host image and that of the private image. A larger $\lambda_3$ can improve the reconstruction performance of the host image and enhance security but may degrade the reconstruction performance of the private image. Conversely, a smaller $\lambda_3$ can enhance the reconstruction performance of the private image while potentially degrading the reconstruction performance of the host image.

\begin{figure*}
    \centering
   \includegraphics[width=0.6\linewidth]{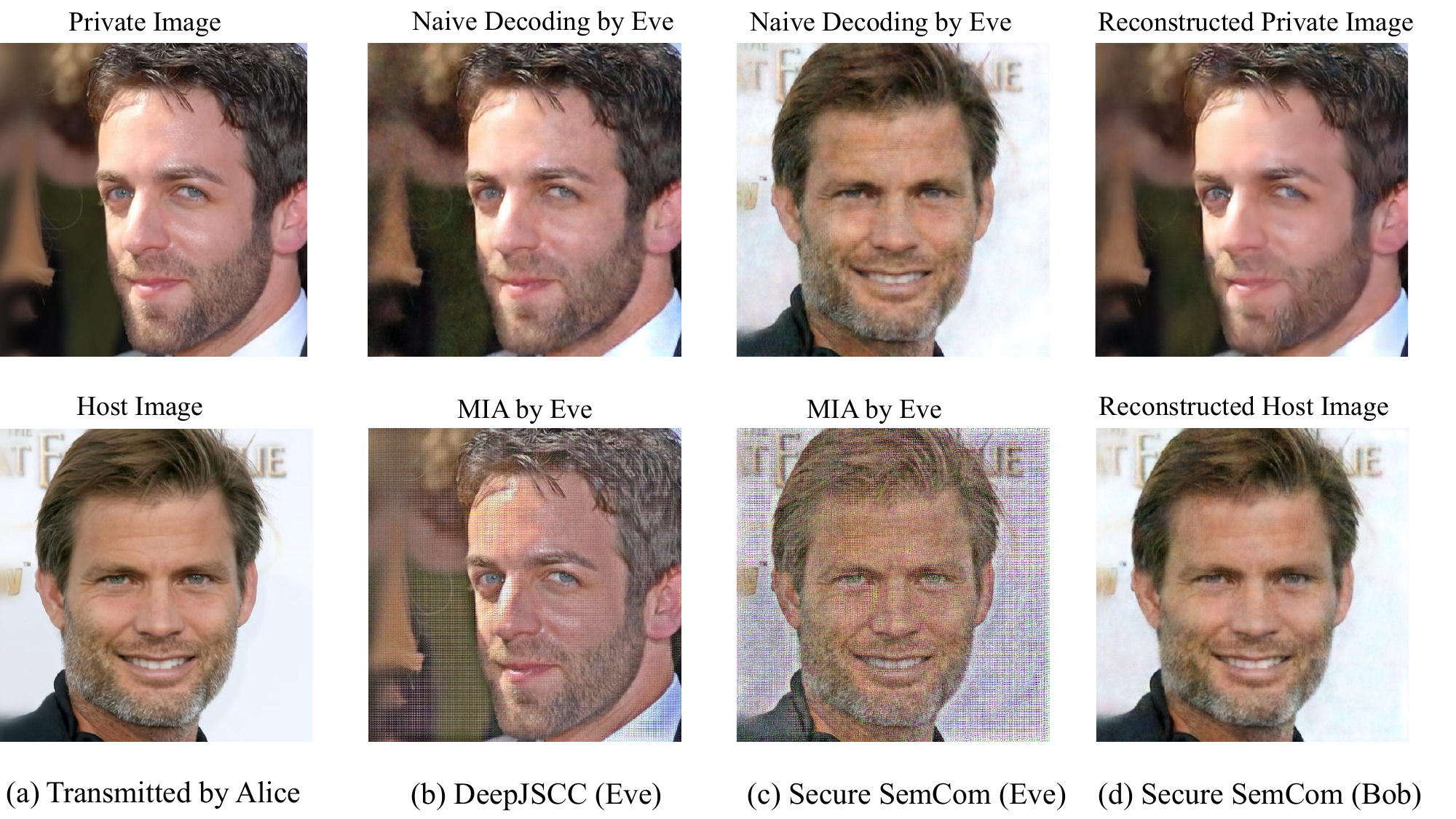}
    \caption{Visual comparison of the host image and private image reconstruction at Bob and Eve, where the SNR of the main link and the eavesdropping link is set to 5 dB. \vspace{-1mm}}
    \label{fig:Visual}

\end{figure*}

 \subsection{Reconstruction performance at Eve}

\autoref{fig:Eve_a} and \autoref{fig:Eve_b} show the PSNR results of the images reconstructed by Eve and the original host and private images, under naive decoding and MIA eavesdropping, respectively, where the SNR of the eavesdropping link varies from 0 to 20 dB. We also present the results of reconstructing private images in DeepJSCC that directly transmit images. From these results, we observe that by the proposed approach images reconstructed by Eve exhibit significant similarity to the original host image, and the PSNR of the private image is less than 10 dB, indicating that Eve can obtain limited information about the private image. In contrast, Eve can easily decode the private image by eavesdropping on the link with DeepJSCC. It is noteworthy that the proposed approach is able to protect the private image against both naive decoding and MIA eavesdropping, which demonstrates its effectiveness in securing image transmission.

We also provide the MS-SSIM and LPIPS results of DeepJSCC and the proposed secure SemCom approach at Eve in the \autoref{tab:Eve}. From the table, we observe that the proposed secure SemCom approach effectively conceals the private image within the host image, and the reconstructed images by Eve are more similar to the host images, which further demonstrates the enhanced security provided by our approach.

  \vspace{-0.5mm}
 \subsection{Visual Comparsion}
\autoref{fig:Visual} presents a visual comparison between the original host image and private image, and the reconstructed images by Bob and Eve, with both the SNRs of the main link and the eavesdropping link set to 5 dB. We observe that the host image and private image reconstructed by the proposed approach closely resemble the original ones, and Eve can only decode the information of the host image. This further demonstrates the effectiveness of the proposed approach in protecting the private image from eavesdroppers.

\vspace{-1mm}
\section{Conclusion}
In this paper, we proposed a novel secure SemCom approach for image transmission, which leverages steganography technology to conceal the latent codes of a private image within that of a host image. Specifically, we employed INN to apply steganography on the channel input signals before transmission, ensuring that the original signals can be precisely reconstructed from the output signals at the legitimate user, while preventing the eavesdropper from decoding any private information. Simulations conducted on CelebAmasked-HQ dataset validated the effectiveness of the proposed approach. The results demonstrated that our proposed approach maintains comparable reconstruction quality of the private image at the legitimate user compared to the scenario without any secure mechanisms. Meanwhile, the eavesdropper is unable to successfully decode the transmitted private image, highlighting the enhanced security provided by our approach.

\bibliographystyle{IEEEtran}
\bibliography{IEEEabrv, references}

\end{document}